\documentclass[preprint,aps,showpacs,amsmath,amssymb]{revtex4}
\usepackage{graphicx} 
\usepackage{dcolumn}
\usepackage{bm}
\begin{document}
\title{Band structure of a two-dimensional (2D) electron gas  in the presence of 2D electric and magnetic modulations and of a perpendicular magnetic field}
\author{X. F. Wang$^\dagger$, P. Vasilopoulos$^\diamond$, and F. M. Peeters$^\star$
\ \\}

\address{$^{\dagger,\diamond}$Concordia University Department of Physics,\\
1455 de Maisonneuve  Ouest, Montr\'{e}al, Qu\'{e}bec, Canada, H3G 1M8\\
\ \\
$^{\star}$Departement Natuurkunde Universiteit Antwerpen\\
(Campus Drie Eiken) Universiteitsplein 1\\ B-2610 Antwerpen, Belgium}
\date{\today}
\begin{abstract}
Two-dimensional (2D) periodic electric modulations of a 2D electron gas split each Landau level into the well-known butterfly-type spectrum
described by a Harper-type equation multiplied by an envelope function. This equation is slightly modified for 2D
magnetic modulations but the spectrum remains qualitatively the same. The same holds if both types of modulations are
present. The modulation strengths do not affect the
structure of the  butterfly-type spectrum,
they only change its scale or its envelope.
The latter is described by the ratio $\alpha$ of the flux quantum $h/e$ to the flux per unit cell.
Exact numerical and approximate analytical results are presented for the energy spectrum
as a function of the magnetic field.
For integer $\alpha$ the internal structure collapses into a band for all cases.
The bandwidth at the Fermi energy depends on
the modulation strength, the electron density, and, when both modulations are present,
on the phase difference between them. In the latter case if the
modulations have a $\pi/2$ phase difference, the bandwidth at the Fermi energy is nearly independent of the magnetic field and the commensurability oscillations of the  diffusive contribution to the resistivity disappear.
\end{abstract}

\maketitle

\section{ INTRODUCTION}
In the last fifteen years the magnetotransport of the two-dimensional electron gas (2DEG),
subjected to periodic potential
modulations, has attracted considerable experimental
and theoretical
attention. For one-dimensional (1D) modulations novel
oscillations of the magnetoresistivity tensor $\rho_{\mu\nu}$
have been
observed \cite{1}, at low magnetic fields $B$, distinctly different
in period and temperature dependence from the usual Shubnikov-de Haas (SdH)
ones observed at higher $B$. These oscillations reflect the commensurability
between two length scales \cite{2}: the cyclotron diameter at the Fermi level
$2R_c = 2\sqrt{2\pi n_e} \ell^2$, where $n_e$ is the electron
density, $\ell$ the magnetic length, and $a$ the period of the potential
modulation.

The situation is similar but less clearcut for 2D electric modulations
from both a theoretical \cite{3,4} and an experimental \cite{5}  point of view. In general, for
2D modulations  a tight-binding treatment shows that each Landau level exhibits the well-known butterfly-type spectrum described by a Harper-type equation. For sinusoidal modulations the energy spectrum resulting from the numerical
solution of this equation shows, when
the energy $E$ is measured in units
of $V_\mu F_n(u_\mu ), \mu=x,y$,
with $V_\mu$ the modulation potential and $F_n(u_\mu )$ the Laguerre polynomial,
a nontrivial structure as a function of
$\alpha = \Phi/\Phi_0$, where $\Phi_0=h/e$ is the flux quantum and $\Phi$ the flux per unit cell:
for $\alpha = p/q$ rational, with $q, p$   integers and relative prime, each Landau level is
split into $p$ subbands \cite{6}.
To our knowledge, despite various efforts \cite{7} there is no conclusive experimental evidence for this
structure yet presumably because the small gaps between these subbands are closed due to disorder
in samples of not exceptionally high mobilities. By neglecting those small gaps we recently
accounted \cite{8} for the observed \cite{9} oscillations in the amplitude of the commensurability
oscillations, as a function of the flux
through the unit cell of the 2D modulation lattice,
that could not be fully explained by earlier
semiclassical theories \cite{3}. The  high resistivity peaks observed for $\alpha$ integer were due to collisional contributions and show up only for short-period  modulations.

In view of recent work on  1D magnetic modulations \cite{10,11}, it is of interest to present
a tight-binding treatment of 2D magnetic modulations in order to see the similarity and the differences
between the former and the latter as well as between 2D electric and magnetic modulations.
So far we are aware only of the work of Ref. 12
which considered only a 2D magnetic modulation and
didn't show any spectra as a function
of $\alpha = q/p$. In addition, we would like to present such a treatment when both types of modulations
are present.
The reason for considering the latter
case is that from experimentally known methods of producing
a {\it magnetic}
modulation one expects that the magnetic \cite{10} or superconducting
stripes \cite{11}, periodically placed on top of a 2DEG, also act like electrical
gates and induce an {\it electric} modulation of the 2DEG. As shown in
our previous work on transport \cite{13}  and the related band structure \cite{14}, the
phase difference between the two 1D modulations can have important consequences and even suppress the {\it Weiss} oscillations; this prediction was later confirmed experimentally \cite{10}.
We show here that similar effects must occur for 2D modulations with a $\pi/2$ phase difference
since the bandwidth at the Fermi energy
is nearly independent of the magnetic field and the commensurability oscillations of the  diffusive contribution to the resistivity disappear. Further, as for 1D modulations \cite{13}, the bandwidth at the Fermi energy for a purely electric modulation oscillates, as a function of the magnetic field, in {\it antiphase} with that for a purely magnetic one. In addition, we show how a small broadening can close the very small gaps in the density of states for $\alpha = q/p$ rational.

In the next section we derive the relevant Harper-type equations and solve them exactly,
for $\alpha$ rational, and analytically, for $\alpha$ integer, to obtain the one-electron eigenfunctions and
eigenvalues. We also present
the density of states and  the bandwidth at the Fermi energy. Various numerical results
are presented in Sec. III and  concluding remarks in Sec. IV.

\section{Formalism}
We consider a 2DEG in the $(x,y)$ plane under a perpendicular magnetic field $B$,
modulated weakly and periodically along the $x$ and $y$
directions, and accompanied by a 2D electric modulation of the same period.
In  Fourier space the total electric and magnetic fields can be expressed, respectively,  by a vector potential
\begin{equation}\label{vp}
\mathbf{A}(\mathbf{r})=\hat{\mathbf{x}}\sum_{\mathbf{g}}
A_{\mathbf{g}}^{x}e^{i\mathbf{g}\cdot r}
+\hat{\mathbf{y}}\left(Bx+\sum_{\mathbf{g}}A_{\mathbf{g}}^{y}
e^{i\mathbf{g}\cdot r}\right)
\end{equation}
and by a scalar potential
\begin{equation}
\phi(\mathbf{r})=\sum_{\mathbf{g}}\phi_{\mathbf{g}}e^{i\mathbf{g}\cdot r}.
\end{equation}
Here $a_x=2\pi/K_x$ and $a_y=2\pi/K_y$ are the lattice periods of the imposed modulations and
$\mathbf{g}=\hat{\mathbf{x}}n_xK_x+\hat{\mathbf{y}}n_yK_y$ is the corresponding reciprocal lattice vector with $n_x, n_y$ integers.
For $\nabla \cdot \mathbf{A=0}$, which can be satisfied in most practical situations,
we have $A_{\mathbf{g}}^{x}g_x+A_{\mathbf{g}}^{y}g_y=0$
and the magnetic field reads
\begin{equation}\label{bm}
\mathbf{B=}\nabla \times \mathbf{A}=\hat{\mathbf{z}}(B
+\sum_{\mathbf{g}}B_{\mathbf{g}}^{z}e^{i\mathbf{g}\cdot r}),
\end{equation}
with $B_{\mathbf{g}}^{z}=ig_xA_{\mathbf{g}}^{y}-ig_yA_{\mathbf{g}}^{x}$ the Fourier component of the magnetic modulation.

Neglecting the spin of the electron and the Zeeman term,
the one-electron
Hamiltonian of the 2DEG reads
\begin{eqnarray}\label{hamt}
\nonumber
H &=&[\mathbf{p}
+e\mathbf{A(r)}]^{2}/2m^{\ast}+V(\mathbf{r}) \\ \nonumber
&=&p_x^{2}/2m^{\ast }+(p_y+eBx)^{2}/2m^{\ast }
+\frac{e}{m^{\ast }}\sum_{\mathbf{g}}e^{i\mathbf{g}\cdot
r}(A_{\mathbf{g}}^{x}p_x+A_{\mathbf{g}}^{y}p_y
+eBxA_{\mathbf{g}}^{y}) \\
&&+\frac{e^{2}}{2m^{\ast}}
\sum_{\mathbf{g}}D_{\mathbf{g}}e^{i\mathbf{g}\cdot
r}+\sum_{\mathbf{g}}V_{\mathbf{g}}e^{i\mathbf{g}\cdot r},
\end{eqnarray}
where $\mathbf{p}$ is the momentum operator,
$D_{\mathbf{g}}=\sum_{\mathbf{g}^{\prime }}
(A_{\mathbf{g}-\mathbf{g}^{\prime
}}^{x}A_{\mathbf{g}^{\prime}}^{x}
+A_{\mathbf{g}-\mathbf{g}^{\prime }}^{y}A_{\mathbf{g}^{\prime
}}^{y})$ is the square term of the vector potential, and
$V_{\mathbf{g}}=-e\phi_{\mathbf{g}}$ is the Fourier transform of
the electric potential.

In the absence of both modulations, i.e., for  $B_{\mathbf{g}}^{z}=0$ and $\phi_{\mathbf{g}}=0$,
the system has an energy spectrum of Landau levels
$E_n=\hbar\omega_c(n+1/2)$. In the Landau gauge, the corresponding wave functions are
\begin{equation}\label{nmag}
\langle \mathbf{r}| n,k_y\rangle\equiv\psi_{n,k_y}=L_y^{-1/2}\exp(ik_yy)\phi_n(x+x_0)
\end{equation}
with $x_0=l^2k_y$, $l^2=\hbar/eB$, and $\phi_n(x)$ the normalized harmonic
oscillator function. Constraining the center of the cyclotron orbit $-x_0$
to be within the sample and using periodic boundary condition along $y$,
gives the degeneracy of the Landau levels as
$L_xL_y/2\pi l^2=N$, with $L_x$ and $L_y$
the sample dimensions.
We choose N to be an integer.

\subsection{Tight-binding treatment}

In the presence of {\it weak} 1D modulation, when coupling between different Landau
levels due to modulation is negligible, the Landau levels are
broadened into bands and their bandwidth oscillates with the uniform field $B$. The eigenstates of the system and its energy spectrum
can be calculated using perturbation
 theory  \cite{13} or  following
a tight-binding treatment \cite{15}.

For 2D modulations, the situation is much more complicated and  in general the eigenstates cannot be the same as those without modulation.
A tight-binding method \cite{155}
used
by Hofstadter \cite{6} to treat
a 2DEG with strong electric modulation but weak perpendicular magnetic
field results in a "butterfly"-type energy spectrum.
Neglecting  coupling between different Landau levels, we can express the new eigenstates as linear combinations of the states without modulation for each Landau level \cite{16}.
In the following, we will show
that using the same method the electron states in the presence
of both electric and magnetic modulations can also be obtained.
Notice, however, that in this method the parameter $\alpha$ is  replaced by its inverse, \cite{macd} i.e., in what follows we will have  $\alpha=\Phi_0/\Phi=q/p$.

For {\it weak} modulations, i.e., for  $B_{\mathbf{g}}^{z}\ll B$
and $V_g \ll \hbar\omega_c$, we neglect the coupling between Landau levels.
Then from Eq. (4) we obtain, for the $n$th Landau level, the matrix elements
\begin{equation}\label{hen}
\left\langle n,k_y^{\prime }\left| H\right| n,k_y\right\rangle
=E_{n}\delta _{k_y^{\prime },k_y}
+\sum_{\mathbf{g}}U_{\mathbf{g}}e^{-il^{2}g_x(2k_y+g_y)/2}\delta
_{k_y^{\prime},k_y+g_y},
\end{equation}
where $U_{\mathbf{g}}$ is the effective potential
\begin{equation}
U_{\mathbf{g}}=V_{\mathbf{g}} F_{n}(Q)
+(e\hbar/m^{\ast})
B_{\mathbf{g}}^{z}G_{n}(Q)+(e^2/2m^{\ast})
D_{\mathbf{g}}F_{n}(Q),
\end{equation}
with $F_{n}(x)=\exp(-x/2)L_{n}(x)$, $G_{n}(x)
=-(\partial /\partial x)
F_{n}(x)$, $Q=l^{2}(g_x^{2}+g_y^{2})/2$, and $L_n(x)$ the Laguerre polynomial of order $n$.

>From Eq. (\ref{hen}) we  see that if only the modulation along the $x$ direction is present, i.e. for $g_y=0$,
there results   only a level  broadening for each state but no coupling between different states.
As noticed earlier \cite{15,9}, the modulation along the $y$ direction, however, introduces a coupling between the states $\left\langle n,k_y\right|$
and $\left\langle n,k_y+K_y\right|$. This leads to the folding of the energy band and
the formation of a Brillouin zone along the $y$ direction.
Then we can assume a wave function of the form:
$\left\| n,\zeta \right\rangle
=\sum_{\lambda }c_{\lambda }(n,\zeta)\left| n,k_y+\lambda
K_y\right\rangle$
with $\lambda=1,2,...M$ for integer $M=Na_y/L_y$. The secular equation
$H\left\| n,\zeta \right\rangle =\mathcal{E}_{n,\zeta}\left\|n,\zeta \right\rangle $
then becomes
\begin{equation}
\sum_{\lambda }\left\{ H_{\lambda ^{\prime
}\lambda }-(\mathcal{E}_{n,\zeta}-E_n)\delta _{\lambda ^{\prime }\lambda }\right\}
c_{\lambda }(n,\zeta)=0,
\label{sec1}
\end{equation}
with $H_{\lambda ^{\prime }\lambda }=
\sum_{\mathbf{g}}\delta
_{\lambda ^{\prime },\lambda
+g_y/K_y}U_{\mathbf{g}}\exp[-il^{2}g_x
\left(2k_y+(\lambda ^{\prime }+\lambda)K_y\right)/2]$.
For sinusoidal modulations and rational values  of $\alpha$ Eq.(8) becomes an extension of Harper's equation, see below.

For a rational value of
\begin{equation}
\alpha =\Phi _{0}/\Phi =2\pi l^{2}/a_xa_y=q/p
\end{equation}
with $q$ and $p$ integers and relative prime,
the $M$-dimensional matrix $H_{\lambda,\lambda ^{\prime }}$ satisfies the periodic condition
$H_{\lambda+p,\lambda ^{\prime}+p}=H_{\lambda,\lambda ^{\prime }}$. Due to the peculiar character of the
harmonic function centered at $x_0=l^2k_y$, this corresponds to an effective spatial periodicity of the Hamilitonian
with period $pl^2K_y=qa_x$ and  Bloch-type extensive states along the $x$ direction emerge
as a general property of the superlattice.
It is convenient \cite{16} to introduce a new Bloch-type basis
$\left\| \left| n,s,k_x,k_y\right\rangle \right.$ with a new quantum number $k_x$,
the momentum of electrons along the $x$ direction,
\begin{equation}
\left\| \left| n,s,k_x,k_y\right\rangle \right. =(N_x)^{1/2}
\sum_{\beta=1}^{N_x}e^{-il^{2}k_x[k_y+(\beta p+s)K_y]}\left|
n,k_y+(\beta p+s)K_y\right\rangle,
\end{equation}
where $N_x=L_x/qa_x$ is assumed an integer by choosing $L_x$.

In this basis the eigenstate of the Hamiltonian reads
\begin{equation}
\left| \left| n,j,k_x,k_y\right\rangle \right\rangle
=\sum_{s=1}^{p}u_{s}(n,j,k_x,k_y)\left\| \left| n,s,k_x,k_y\right\rangle \right..
\end{equation}
Since $u_s(...)$ obeys the relation $u_{s+p}(n,j,k_x,k_y)=u_{s}(n,j,k_x,k_y)$, it is a periodic function of $s$ and
the $M \times M$ Hamiltonian matrix is reduced into a $p\times p$ one with  elements
\begin{equation}
[h_{n}^{(p)}(k)]_{s^{\prime },s}=\sum_{\mathbf{g}}
\tilde{\delta}_{[g_y/K_y],s^{\prime }-s}^{(p)}
U_{n,\mathbf{g}}
\exp[-il^{2}(g_xk_y-g_yk_x+sg_xK_y+g_xg_y/2],
\end{equation}
and
\begin{equation}
\tilde{\delta} _{\lbrack g_y/K_y],s^{\prime }-s}^{(p)}=\left\{
\begin{tabular}{l}
1 if $[g_y/K_y]=s^{\prime }-s$ or $s^{\prime }-s+p$ \\
0 otherwise
\end{tabular}
\right.,
\end{equation}
$[g_y/K_y]$ is $g_y/K_y$ modulo $p$,
$k_x \in [0,2\pi/qa_x]$, $k_y\in [0,2\pi/a_y]$, and $s=1,2,\cdots,p$.
Each Landau level is split into $p$ minibands with energy denoted by $\mathcal{E}^{(p)}_{n,s}(k_x,k_y)$.

In this paper we will discuss in detail the energy spectrum
and the density of states (DOS)
of a 2DEG in a constant magnetic field $B$ with a weak magnetic modulation
$\mathbf{B}=[em^{\ast }\omega _{x}\cos
(K_{x}x)+em^{\ast }\omega _{y}\cos (K_yy)]\hat{\mathbf{z}} $ and a weak electric
modulation of the potential $V=V_{x}\cos (K_{x}x+\varphi
_{x})+V_{y}\cos (K_{y}y+\varphi _y)$ with $\varphi _x, \varphi _y$ the phase differences between
the respective components of the potential.
The corresponding effective potential is found to be
\begin{eqnarray}
\nonumber
U_{n} &=&\sum_{\mathbf{g}}U_{n,\mathbf{g}}e^{i\mathbf{g}\cdot r}
=V_{x}F_{n}(u_x)\cos (K_{x}x+\varphi _{x})+V_{y}F_{n}(u_y)\cos
(K_{y}y+\varphi _{y}) \\ \nonumber
 &&+\hbar \omega _{x}G_{n}(u_x)\cos
(K_{x}x)+\hbar \omega _{y}G_{n}(u_y)\cos (K_{y}x)\\ 
&&+\epsilon_x [1-F_{n}(4u_x)\cos (2K_{x}x)]+
\epsilon_y
[1-F_{n}(4u_y)\cos (2K_{y}y)],
\end{eqnarray}
with $u_\mu=l^{2}K_\mu^{2}/2$
and $\epsilon_\mu=m^{\ast}\omega_\mu^2/4K_\mu^2$, $\mu=x,y$.
With $\mathcal{E}^{(p)}_{n,s}(k_x,k_y)\equiv \mathcal{E}_{n,\zeta}$ the corresponding secular equation, Eq. (\ref{sec1}), becomes
\begin{eqnarray}
\nonumber
&&-\frac{\epsilon_y}{2} F_{n}(4u_y)[C_{\lambda -2}+C_{\lambda +2}] +\left\{ \epsilon_x+\epsilon_y
-\epsilon_x F_{n}(4u_x)\cos [2l^{2}K_x(k_y+\lambda K_y)]\right.\\ \nonumber
&&\left.+\hbar \omega
_{x}G_{n}(u_x)\cos [l^{2}K_x(k_y+\lambda K_y)]+V_{x}F_{n}(u_x)\cos
[l^{2}K_x(k_y+\lambda K_y)-\varphi _{x}]-(\mathcal{E}_{n,\zeta}-E_n)\right\}
C_{\lambda } \\
&&+\frac{1}{2}[\hbar \omega _{y}G_{n}(u_y)+V_{y}F_{n}(u_y)e^{i
\varphi _{y}}]C_{\lambda -1}+\frac{1}{2}[\hbar \omega _{y}G_{n}(u_y)+
V_{y}F_{n}(u_y)e^{-i\varphi _{y}}]C_{\lambda +1}=0.
\end{eqnarray}

For $\alpha=2l^{2}K_x K_y=q/p$
rational, with $p,q$ integers and relative prime, the exact
solution of this difference equation is obtained numerically
by rewriting it in the  form of a matrix \cite{6}, \cite{16},  of dimension $p$, and solving the resulting eigenvalue
problem with the details given above. It
will be shown below.

For a pure electric modulation Eq. (15) holds with the terms $\propto\epsilon_\mu$ and
$\propto\omega_\mu$ set equal to zero while  for a pure magnetic modulation
the terms $\propto V_\mu$ vanish; in either case we set $\phi_\mu=0$.  Comparing these two cases, we see that Eq. (15) should lead to the same overall spectrum if $\hbar\omega_\mu G_{n}(u_\mu)=V_\mu F_{n}(u_\mu)$ provided we neglect the very small quadratic terms $\propto\epsilon_\mu$. Examples of these spectra will be shown in Sec. III.

\subsection{Results for $\alpha$ integer}

For $\alpha=2l^{2}K_x K_y$ integer the dependence of  the quantity $\{...\}$ 
in Eq. (15) on $\lambda$ drops out since
$\lambda$ is an integer. Then Eq. (15) admits the exponential solutions $C_\lambda = C_0
\exp(i\xi \lambda)$;
the resulting energy eigenvalue is
\begin{eqnarray}
\nonumber
\mathcal{E}_{n}^{(1)}(k_x,k_y) &=& E_{n}+\epsilon_x
[1-F_{n}(4u_x)\cos 2\eta
]+\epsilon_y
[1-F_{n}(4u_y)\cos 2\xi]\\ \nonumber
&&+\hbar \omega _{x}G_{n}(u_x)\cos \eta
+\hbar \omega
_{y}G_{n}(u_y)\cos \xi\\
&&+V_{x}F_{n}(u_x)\cos
(\eta -\varphi _{x})+V_{y}F_{n}(u_y)\cos
(\xi+\varphi _{y}),
\end{eqnarray}
where $\xi =  k_x\ell^2 K_y$ and $\eta =  k_y\ell^2 K_x$. This is a simple sinusoidal band as when only one modulation is present \cite{8}.
The corresponding eigenvectors  are labeled with the additional quantum number $\xi$
$(0 \leq \xi \leq 2\pi )$: $\mid \psi_{nk_y\xi} > =C_0 \sum_p \exp(i\xi \lambda)
|n,k_y + \lambda K_y>$. The orthonormality condition
gives $\xi =k_x\ell^2 K_y$ and $C_0 = \ell(K_y/L_x)^{1/2}$ by normalization.

If we set $V_{x}=V_{y}=0$ in Eq. (16), we see that the only difference from the case of a pure electric 2D modulation is the presence of the terms
$\epsilon_\mu$, which are proportional to $ \omega _{\mu}^2$, on the
first line of Eq. (16); however, these terms are very weak compared to  those on the 2nd and 3rd line $\propto \omega _{\mu}$. Further, it is instructive to contrast the spectrum, given by Eq. (16), for a pure electric 2D modulation ($\hbar\omega_\mu=0$) with that for a pure magnetic 2D modulation ($V_\mu=0$) as done  for Eq. (15). We then see explicitly
that the two spectra are the same provided $\hbar\omega_\mu G_{n}(u_\mu)=V_\mu F_{n}(u_\mu)$. This means that the  results of a pure electric modulation can be mimicked by those of a pure magnetic one and vice versa. Notice, however, that this equivalence does not hold for those values of  $u_\mu$ for which  the factors $G_{n}(u_\mu)$ or $F_{n}(u_\mu)$ vanish since they are not the same in both cases. The same conclusion was  reached in Ref. 12 with the corresponding spectrum evaluated by first-order perturbation theory.

In the absence of modulations we have the discrete Landau levels
$E_n$. In their presence  Eq. (16) shows that these levels are broadened into the bands $\mathcal{E}_{n}^{(1)}(k_x,k_y)$
 with  explicit sinusoidal dependence on $k_x$ and $k_y$. 
If we neglect the very small terms $\propto \epsilon_\mu$ on the first line and take
$\varphi _{x}=\varphi _{y}=0$ for simplicity, their bandwidth is given by
$\Delta E_n(u_x,u_y)=2[V_xF_{n}(u_x)+V_yF_{n}(u_y)+\hbar \omega_x G_{n}(u_x)+\hbar \omega_y G_{n}(u_y)]$
and it oscillates with the field $B$.
The related velocity components ($v_\mu=(1/\hbar)(\partial \mathcal{E}_{n}^{(1)}(k_y,k_x)/\partial k_\mu
), \mu=x,y)$ resulting from Eq. (16) are
\begin{equation}
v_x=
-(l^{2}K_y/\hbar)\left[V_{y}F_{n}(u_y)
\sin (\xi +\varphi_y)
+\hbar\omega _{y}G_{n}(u_y)\sin \xi
-2\epsilon_yF_{n}(4u_y)\sin
2\xi\right],
\end{equation}
\begin{equation}
v_y=-(l^{2}K_x/\hbar)\left[V_{x}F_{n}(u_x)\sin (\eta -\varphi_x)
+\hbar\omega _{x}G_{n}(u_x)\sin\eta
-2\epsilon_xF_{n}(4u_x)\sin
2\eta \right].
\end{equation}
The broadening of the Landau levels into bands that gives rise to these velocity components has important consequences
for transport \cite{4} especially when the fine structure of the exact energy spectrum is not resolved due to disorder \cite{8}.
The main consequence is  a non vanishing {\it diffusive} contribution to the conductivity which is absent when the modulations are not present.

In the following we will show that we can use the expression of the energy spectrum, Eq. (16),
and the velocity operators, Eqs. (17) and (18), for $\alpha$ integer to approximately estimate the physical properties
of the 2DEG system for any value of $\alpha$.

\subsection{The density of states}

The energy spectrum given by $\mathcal{E}^{(p)}_{n,s}(k_x,k_y)$ is qualitatively
different from the unmodulated spectrum, given by $E_n$, and
from the corresponding 1D modulation spectrum given by
$E_n + F_n(u_x)\cos K_x x_0$. These differences are also
reflected in the density of states (DOS) defined by
\begin{equation}
D(E) = 2\sum_{n,s,k_x,k_y} \delta(E - \mathcal{E}^{(p)}_{n,s}(k_x,k_y)).
\end{equation}
For a Lorenzian broadening of width $\Gamma$, the $\delta$ function in the above equation should be replaced
by $(1/\pi)\Gamma/[\Gamma^2+(E - \mathcal{E}^{(p)}_{n,s}(k_x,k_y))^2]$.

The
result  for $\alpha$ integer is obtained by replacing $\mathcal{E}^{(p)}_{n,s}(k_x,k_y)$ by $\mathcal{E}^{(1)}_{n,s}(k_x,k_y)$ in Eq. (19).
For a 2D modulation with rectangular symmetry,
corresponding to Eq. (16), the DOS becomes
\begin{eqnarray}
D(E ) = &&D_0 \sum_{n=0}^{\infty} \int^{2\pi}_0 d\xi
\{ \left[ V_xF_n(u_x)+\hbar\omega_xG_n(u_x)\right]^2\\ \nonumber
&&-\left[E - E_n - \{V_y F_n(u_y)-\hbar\omega_yG_n(u_y)\}\cos\xi\right]^2 \}^{-1/2},
\end{eqnarray}
where $D_0 = L_y L_x/\pi^3\ell^2$.
The radicant in Eq. (20)  must be positive.

\section{Results and discussion}

\subsection { In-phase, antiphase modulations}

The results presented so far  are valid for rectangular modulations. However, as the various expressions and results become simpler for square modulations, we will consider only the latter.
For a square modulation where
$K_x=K_y=K,\varphi_x=\varphi _y=\varphi,
V_x=V_y=V,\omega_x=\omega _y=\omega$,
the secular equation (\ref{sec1}), without the square terms  $\propto \epsilon_\mu$ due to the magnetic modulation,
is Harper's equation
with $\varphi=0$ if the  modulations are in phase and  $\varphi=\pi$ if they are in antiphase;
in these cases Eq. (8)
takes the form
\begin{equation}
C_{\lambda -1}/2+C_{\lambda +1}/2+C_{\lambda }\cos (\lambda
l^{2}K^{2}+Kx_{0})-C_{\lambda }\epsilon _{n,\zeta }=0,
\end{equation}
where $\epsilon _{n,\zeta }=(\mathcal{E}_{n,\zeta}-E_n)/[VF_{n}(u)\pm\hbar \omega G_{n}(u)]$.
The resulting energy spectrum is the "butterfly" spectrum $\epsilon^{(p)}_{n,s}(k_x,k_y)\in [-2,2]$
modulated by an envelope
function determined by the modulations
\begin{equation}
\label{eng}
\mathcal{E}^{(p)}_{n,s}(k_x,k_y)=E_{n}+[VF_{n}(u)\pm\hbar \omega G_{n}(u)]
\epsilon^{(p)}_{n,s}(k_x,k_y),
\end{equation}
where the upper (lower) sign is for  in-phase (antiphase) modulations.

The  spectrum for $\alpha$ integer and the corresponding
wave function were used previously \cite{8} as an approximation,  for all values of the magnetic field, and successfully described magnetotransport of a 2DEG in the presence
of a 2D electric modulation in the case that the fine structure of the exact
spectrum is not resolved.
This approximation implies the replacement of
$\epsilon^{(p)}_{n,s}(k_x,k_y)$ by
$\epsilon^{(1)}_{n,1}(k_x,k_y)=\cos\xi
+\cos\eta$ in Eq. (\ref{eng}),
which becomes
\begin{equation}
\mathcal{E}^{(p)}_{n,s}(k_x,k_y)\approx E_{n}+[VF_{n}(u)\pm\hbar \omega G_{n}(u)](\cos\xi
+\cos\eta).
\end{equation}
In the following the resulting spectrum will be referred to as approximate band structure or
approximate spectrum.

The edge of the approximate band is the envelope of the exact energy spectrum.
Employing the asymptotic expressions of  the
Laguerre polynomials for $n\gg 1$,
the corresponding bandwidth is estimated as
\begin{equation}
\Delta E_n(u) 
\simeq 4
(\pi^2nu)^{-1/4}
\tilde{V}(1+\tilde{\delta}^2)^{1/2}
\left|\sin(2\sqrt{nu}-\pi/4+\beta)\right|,
\end{equation}
with $\tilde{\delta}=(\hbar\omega/\tilde{V})\sqrt{n/u}$, $\tilde{V}=V\pm\hbar\omega/4u$ and $\beta=\arctan[\pm(\tilde{V}/\hbar\omega)\sqrt{u/n}]$,
so that the flat band condition is $\beta\approx m\pi+\pi/4-2\sqrt{nu}$ where $n$ is the Landau index and $u=K^2l^2/2$.
For $1/4u\ll \sqrt{n/u}$ or $\alpha \gg 16n\pi$, which is valid for typical parameters discussed in this paper,
we can further approximate $\tilde{V}$ with $V$ and $\tilde{\delta}$ with $\delta=(\hbar\omega/V)\sqrt{n/u}$.
Because $\beta=\pi$ for a  pure electric modulation
and $\beta=0$ for a pure magnetic modulation, the two  modulations are phase shifted and their interplay will influence the results of transport experiments.

\begin{figure}[tpb]
\includegraphics*[width=110mm,height=150mm]{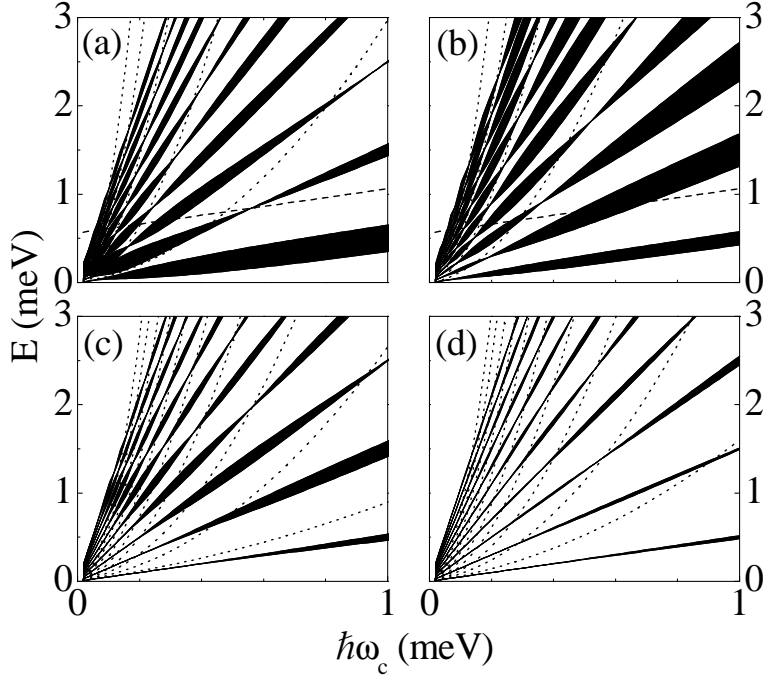}
\vspace{-2.5cm}
\caption{Approximate band structure, for $n=0,...10$, as a function of $\hbar\omega_c$ for
square electric ((a) and (c)) or magnetic ((b) and (d))
modulations of long period ($a=200$nm), in (a) and (b), and short period ($a=80$nm) in (c) and (d).
The crossings of the dotted curves with the Landau levels give the asymptotic flat-band
positions and the dashed line, in panels (a) and (b), is the $\delta=1$ curve.}
\label{fig1}
\end{figure}
In Fig. 1 we plot the envelope function of the energy spectrum for
Landau levels $n=0$ to $n=10$ as a function of the cyclotron energy
$\hbar\omega_c$.  For the same modulation strength
$V=\hbar\omega_0=0.1meV$ the ratio between the band broadenings
is determined by the value of $\delta$, cf. Eq. (24). For a period $a=200$ nm, cf. Figs. 1(a) and (b),
the curve $\delta=1$ is located near $E\approx 0.8meV$ and
the broadening due to magnetic modulation is generally larger than that
due to the electric one. Decreasing the  period $a$, the
$\delta=1$ curve is shifted to a higher energy, higher than 3meV
for $a=80$nm. This is the reason why the magnetic band broadening is
narrower than the electric one for energies lower than 3meV as shown in
Figs. 1(c) and (d).
\begin{figure}[tpb]

\includegraphics*[width=110mm,height=150mm]{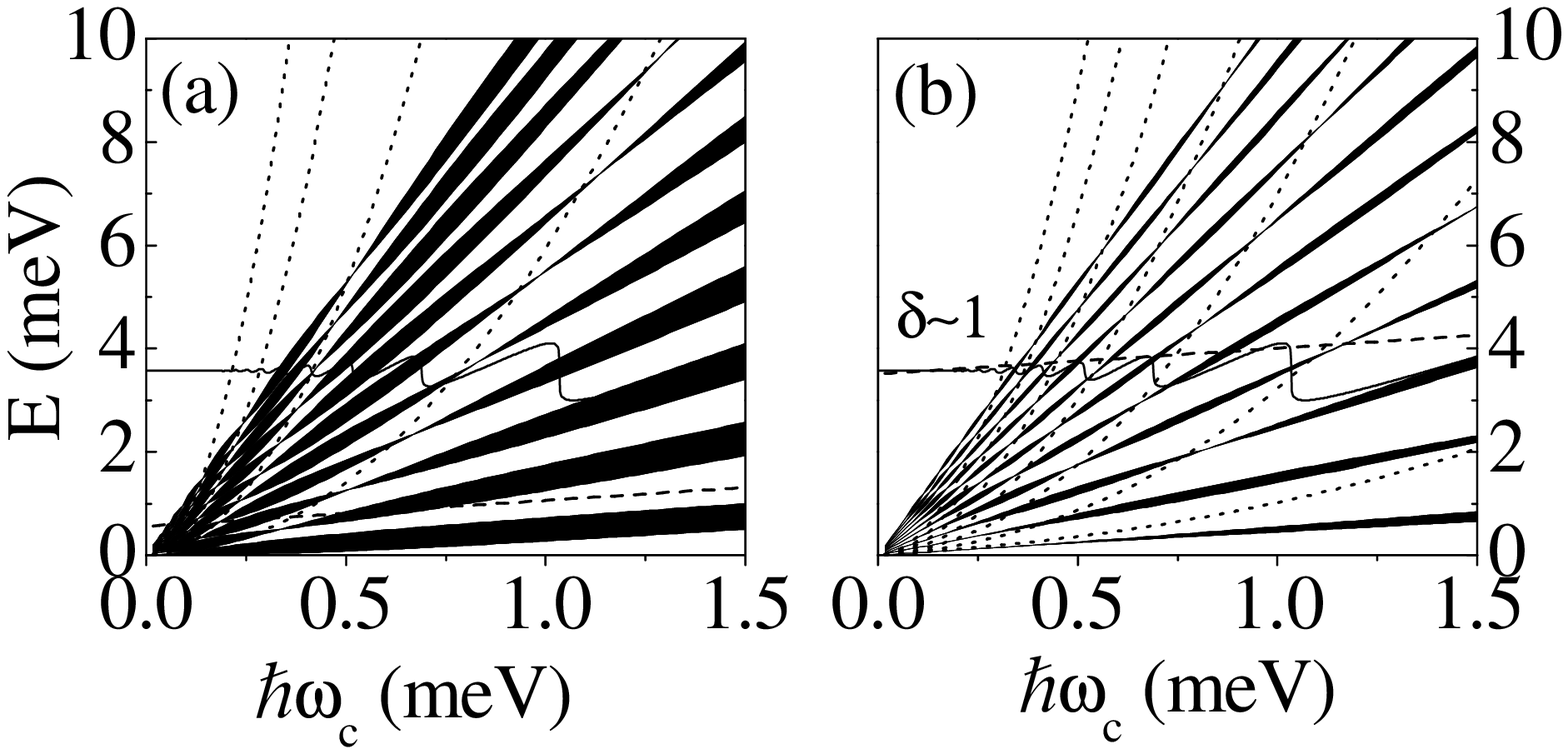}
\vspace{-4cm}
\caption{The same as in Fig.1
 for both electric and magnetic in-phase
modulations of long (a) and short (b) period.
The thin solid curves show the Fermi energy for temperature $T=1$K and electron
density $n_e=10^{11}$cm$^{-2}$.}
\label{fig2}
\end{figure}

\begin{figure}[tpb]
\includegraphics*[width=110mm,height=150mm]{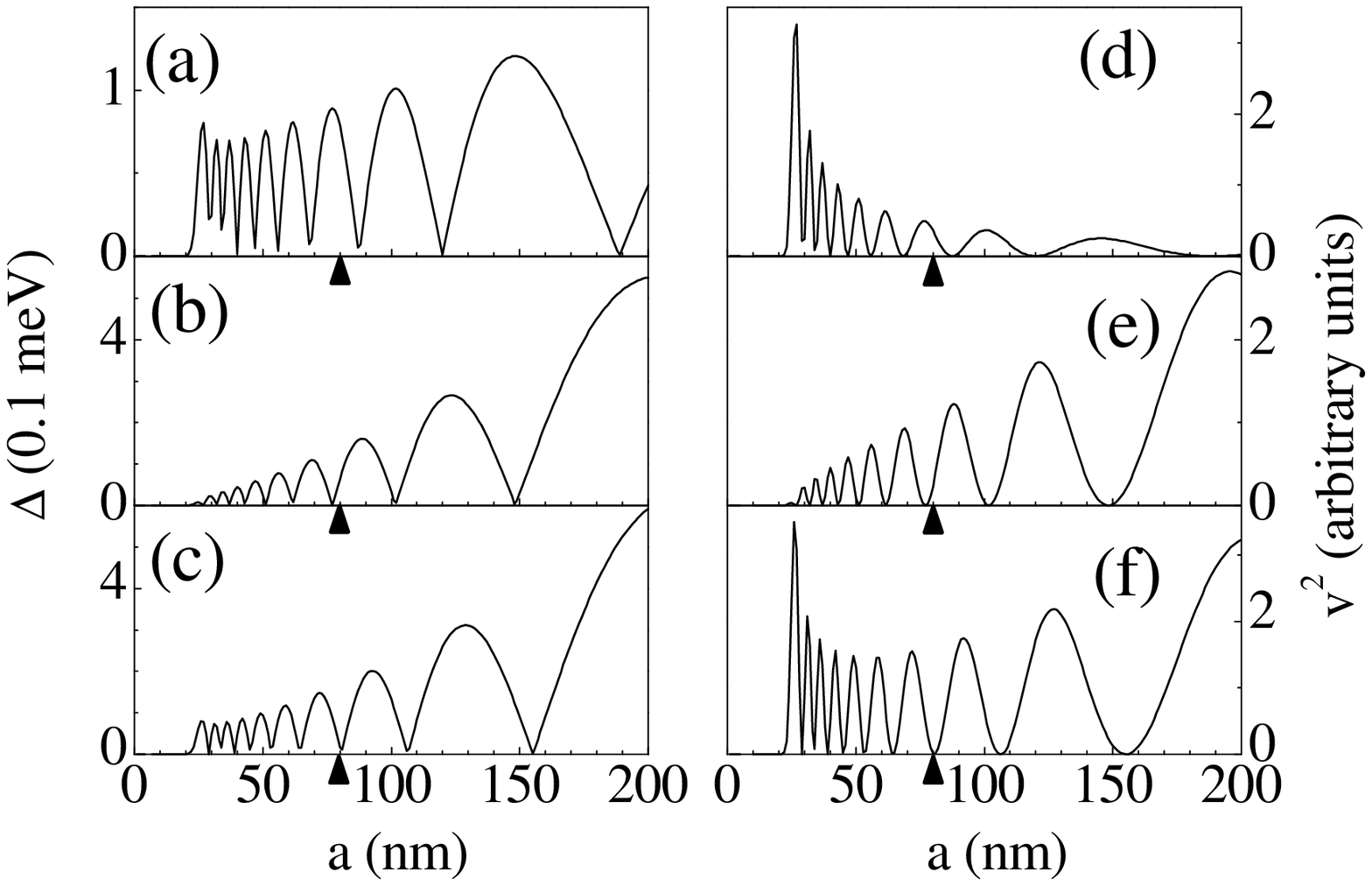}
\vspace{-3.5cm}
\caption{Left panels: the approximate bandwidth $\Delta$ of the 10th Landau
level as a function of the modulation period $a$
at magnetic field $B=0.5$ T for modulation strengths (a)$V=0.1$ meV and $\hbar\omega=0$, (b) $V=0$ and
$\hbar\omega=0.1$ meV, and (c) $V=\hbar\omega=0.1$ meV.
Right panels:
the constant term of the square of the velocity in arbitrary units,
given by Eq. (17) or Eq. (18), vs the modulation period $a$ at magnetic field $B=0.5$. Panels (d), (e), (g) correspond to panels (a), (b), (c), respectively. The two periods used in Figs. 1 and 2 are $80$ nm,
indicated by the triangle, and $200$ nm.}
\label{fig3}
\end{figure}

Figure 2 illustrates the envelope function as in Fig. 1 with both
modulations having the same strength
$V=\hbar\omega_0=0.1$ meV and being in phase. In such a case,
the modulation of longer period results in a broader bandwidth.
This is opposite to that obtained from the band broadening in a common superlattice
and results from the presence of the strong perpendicular magnetic field.
In a common superlattice the miniband appears as a result of electron tunneling
between adjacent quantum wells. The miniband width
decreases when the superlattice period increases since the tunnel coupling weakens if the other parameters remain the same. However, here the level broadening stems mainly from the perturbative correction to the Landau energy of each state by the modulation.
In Fig. 3 we show the resulting bandwidth of the 10th Landau level as a function of the modulation period for fixed magnetic field $B=0.5$ T and the strength parameters shown in the caption. We see that on the average the width increases when the period increases, say by a factor of $s$ provided $s$ is such that   $sa$ is not close to the point where the bandwidth vanishes.
On the right panels of Fig. 3 we show the constant ($\xi$- or $\eta$-indepedent) term of the square of the velocity, given by Eq. (17) or Eq. (18), vs the modulation period $a$ for the same magnetic field. If we neglect the very small terms $\propto\epsilon_\mu$, this term is equal to $[(\ell^2K_y/\hbar)V_{y}F_{n}(u_y)+\hbar \omega _{y}K_yG_{n}(u_y)]^2/2$.  This term gives by far the dominant contribution to the diffusive conductivity, given by the standard expression $\sigma^{dif}_{\mu\nu} (0) = (\beta e^2 / \Omega) \sum_\zeta f_\zeta (1 - f_\zeta)\tau (E_\zeta)v^\zeta_\mu v^\zeta_\nu$, where $f$ is the Fermi-Dirac function,  $v^\zeta_\mu$ the velocity, and $\tau$ the relaxation time \cite{8} pertinent to a state $\zeta\equiv \{k_x, k_y, n\}$. With the integrals over $k_x$ and $k_y$ giving, for this term, just a constant prefactor near the Fermi level, we see that this contribution is mainly determined by the bandwidth. By contrasting the left and right panels one can see that this on-the-average increase of the bandwidth with the period is mainly due to the behavior of the contribution of  the magnetic modulation to it.
This behavior of the bandwidth and the antiphase between panels (a) and (b) on the left, and panels (d) and (c) on the right, are directly related to those of the Laguerre polynomials and their derivatives that appear in the factors $F_n$ and $G_n$.

\begin{figure}[tpb]
\includegraphics*[width=120mm,height=130mm]{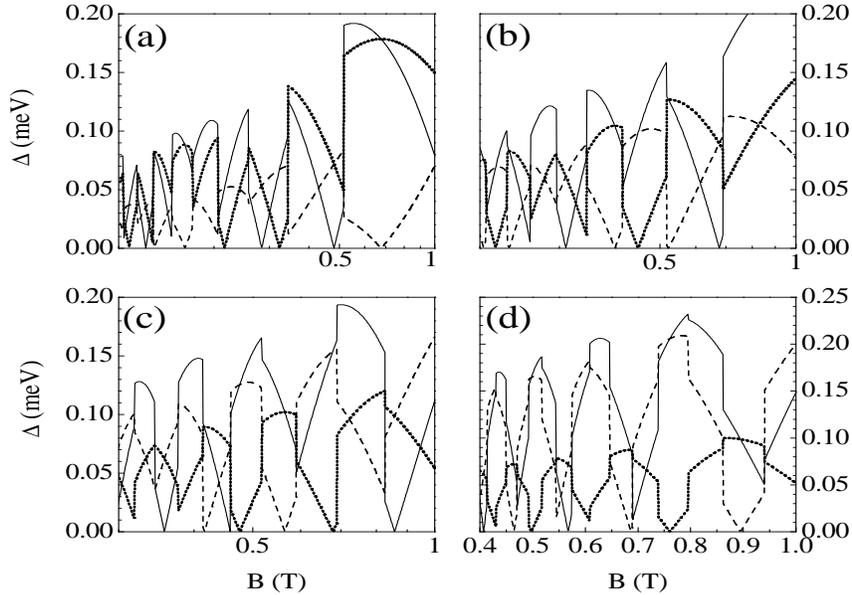}
\vspace{-2cm}
\caption{The approximate bandwidth of the Landau level closest to the
Fermi level as a function of the magnetic field for different electron
densities $n_e=0.5$ (a), 1.0 (b), 2.0 (c), and $5.0\times
10^{11}$cm$^{-2}$ (d). The solid curves denote broadening with both
 modulations present and of the same strength. The dotted
(dashed) curves show the broadening when only the electric (magnetic)
modulation is present.}
\label{fig4}
\end{figure}

If both modulations are present, their relative
contribution to the energy broadening  is determined by the value of $\delta$. The magnetotransport behavior of a 2DEG  is directly related to the level  broadening at the Fermi energy.  Because the electric modulation dominates  the low-energy broadening and
the magnetic modulation dominates the  high-energy one, the
transition from one  energy range to another
can be observed if we shift the Fermi energy through these ranges by changing the electron density.
In Fig. 4 we plot the
bandwidth at the Fermi energy for a period $a=80$nm as a
function of the perpendicular magnetic field $B$. The bandwidth
oscillations,  resulting when
only the electric modulation is present, are in antiphase with those resulting when only the magnetic one  is present.
This should show up in
the commensurability oscillations of the magnetoresistance
as it does for 1D modulations \cite{13}. If
both modulations have the same strength, the
bandwidth oscillations (solid curves) are similar to those due to
the electric one at the lower electron density (Fig. 4(a)) and similar to those due to
the magnetic one at the higher electron density (Fig. 4(d)).
The non smooth behavior of the bandwidth, when the field $B$ is varied, reflects that of the Fermi level,
as shown, for instance, in Fig. 2. Upon changing the electron density the Fermi level crosses different Landau levels in a non smooth manner, cf.  Fig. 2. For higher densities and/or longer periods more Landau levels are occupied and the oscillations are much smoother.
\begin{figure}[tpb]
\includegraphics*[width=110mm,height=150mm]{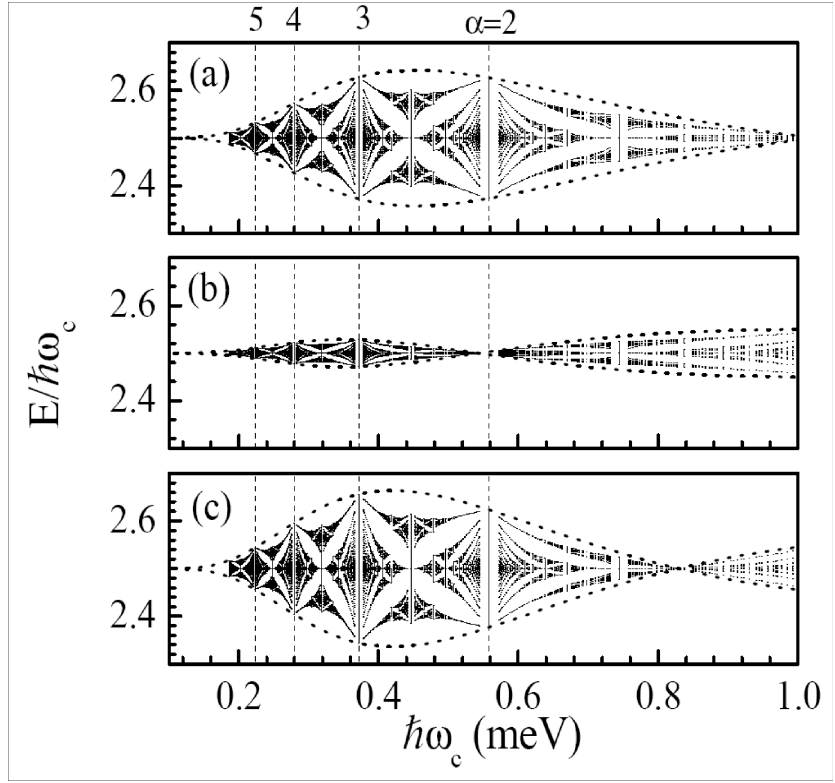}
\vspace{-1.3cm}
\caption{Exact and approximate ($\alpha$ integer) spectrum, in units of $\hbar\omega_c$, vs the cyclotron energy
$\hbar\omega_c$
for the $n=2$ Landau level and a period  $a=80$ nm. The upper panel is for an electric modulation, the middle one
for a magnetic one, and the bottom one for both modulations present. Notice the continuous bands for $\alpha$ integer: in this case the exact, numerically obtained bandwidth coincides with  the one obtained analytically.}
\label{fig5}
\end{figure}

In Fig. 5 we plot
the exact energy spectrum of the second Landau level,  in units of $\hbar\omega_c$, as a function of $\hbar\omega_c$
for (a) only an electric modulation, (b) only a magnetic modulation, and (c) both electric and magnetic modulations
all of period $a=80$nm and strength $V=\hbar\omega=0.1$meV. The positions of  $\alpha=q/p=2,3,4,5$ are marked by the dashed vertical lines. The exact spectrum is the same as Hofstadter's
spectrum modulated by the envelope function (dotted curves in Fig. 5). For a rational value of $\alpha=q/p$ the spectrum is composed of $p$ minibands which touch each other
for $\alpha$  half-integer or $p=2$.
For $\alpha$ integer  the spectrum is the sinusoidal band given by Eq. (23). The envelope function, given by the dotted curves, is the band edge of the $\alpha$ integer spectrum [Eq. (23)] used here for all values of $\alpha$.
A negligible shift and the modification of the spectrum, due to the
terms  $\propto\epsilon_\mu$, are not shown in the figures.

In Fig. 6 we show the exact and approximate DOS for in-phase modulations and various values of $\alpha$. The other parameters are specified in the caption. The results coincide, as they should, for $\alpha=2$. Notice the horizontal scale and how a small level broadening $\Gamma$ closes the gaps of the exact DOS in panels (b)-(d).
\begin{figure}[tpb]
\includegraphics*[width=120mm,height=110mm]{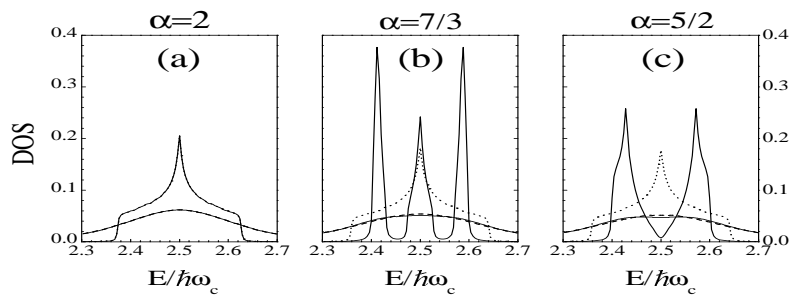}
\vspace{-2.3cm}
\caption{Broadened DOS for the exact (thin solid curves: $\Gamma=0.01$K,  thick curves: $\Gamma=0.5$K)
and the $\alpha$ integer spectrum approximation (dotted curves: $\Gamma=0.01$K, dashed curves: $\Gamma=0.5$K).
The magnetic field is such that $\alpha=2$ in (a), $\alpha=7/3$ in (b), $\alpha=5/2$ in (c). $V=\hbar\omega_0=0.1meV$}
\label{fig6}
\end{figure}

\subsection{Out-of-phase modulations}
For arbitrary phase difference between the two
modulations, the secular equation cannot be written
simply as Harper's equation
but as a  more general
one, cf. Eq. (15).
However, the energy spectrum for $\alpha$ integer is still
the envelope of the exact spectrum and the asymptotic bandwidth is
\begin{equation}
\Delta E_n(u)=4V (\pi^2nu)^{-1/4}
\left[1+(\delta^2-1)\sin^2(2\sqrt{nu}- \pi/4)-\delta\cos\phi\cos(4\sqrt{nu})\right]^{1/2},
\end{equation}
where $\delta=(\hbar\omega/V)(n/u)^{1/2}$. At the Fermi energy we have
$\delta(E_f)=ak_F\hbar\omega/(2\pi V)$.
If $\delta=1$ and $\varphi=\pi/2$, we get a nearly constant bandwidth
$4V(\pi^2nu)^{-1/4}$.

In Fig. 7 we plot the energy of the $n=0,...,10$ Landau levels as a function of the
magnetic field for  equal-strength ($0.1$ meV) modulations, of period $80$ nm, that are out of phase by $\pi/2$.
The dotted curve, i.e. $\delta=1$, shows a constant bandwidth that, as explained above, leads  to a washout of the commensurability oscillations. The thin solid curve shows the  Fermi level at $T=5$ K.
\begin{figure}[tpb]
\includegraphics*[width=110mm,height=150mm]{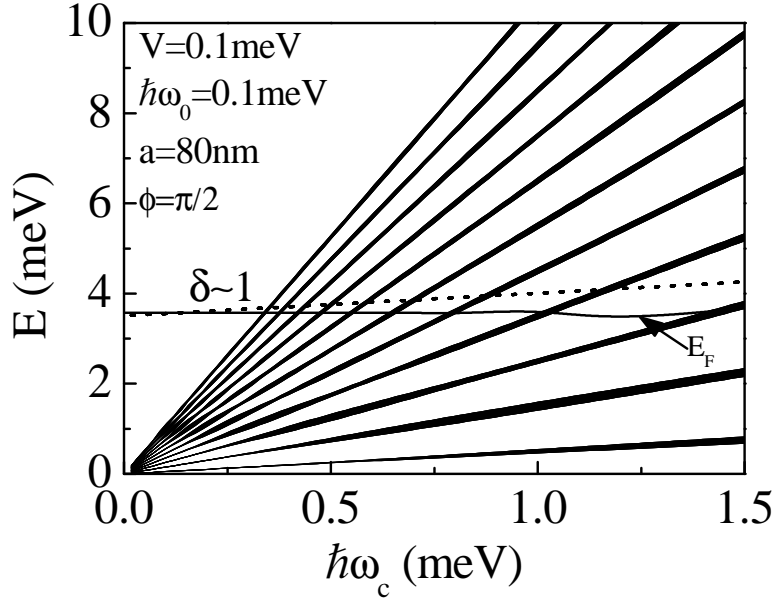}
\vspace{-2cm}
\caption{Band structure with both modulations present and
phase-shifted by $\phi=\pi/2$.
The Fermi level at temperature $T=5$ K is also indicated by the thin solid curve.
The $\delta\sim 1$ curve shows a constant bandwidth leading to a washout of the commensurability
oscillations of the diffusive contributions to the resistivity.}
\label{fig7}
\end{figure}

\begin{figure}[tpb]
\includegraphics*[width=110mm,height=110mm]{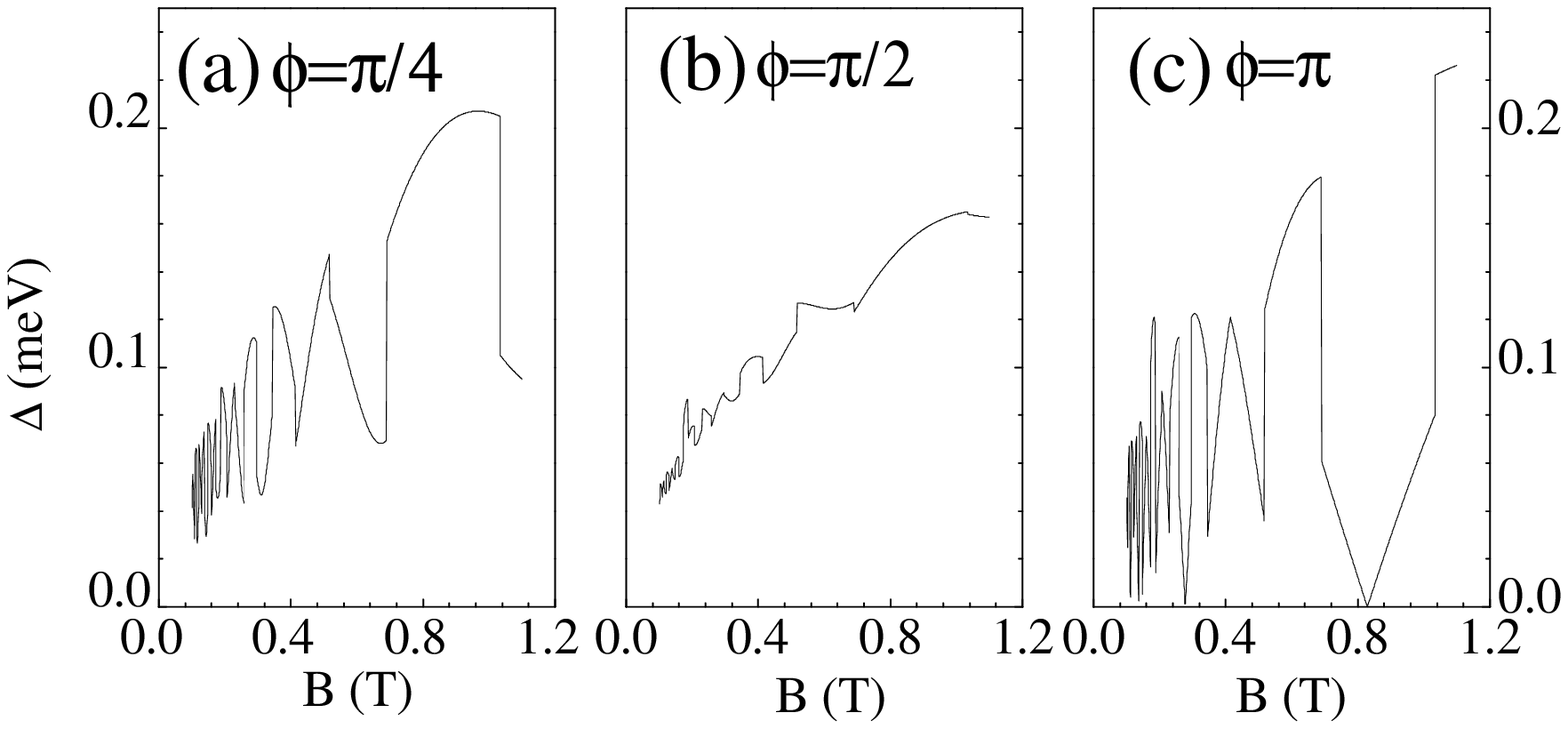}
\vspace{-2cm}
\caption{Bandwidth at the Fermi energy as a function of B for different
phases $\phi=\pi/4$ (a), $\phi=\pi/2$ (b), and
$\phi=\pi$ (c) between the two modulations.}
\label{fig8}
\end{figure}

The amplitude of the bandwidth oscillations varies with the phase
difference between the two modulations and/or with their relative strength.
In Fig. 8 we plot the bandwidth  for equal  modulation strengths,
$V=\hbar\omega=0.1$ meV,  and different
phases. As  Fig. 8(b) shows,
the amplitude can be greatly decreased
when $\phi=\pi/2$ and this implies a  washout of
the commensurability oscillations.
Different from the in-phase or antiphase case (Fig. 4), we notice that the bandwidth increases more
rapidly with
magnetic field and that it
does not reach zero at specific values of $B$.
In Fig. 9 we show again the bandwidth for modulations phase shifted by $\pi/2$ but with different relative strengths apart from panel (c), in
which the two strengths are equal, and   the amplitude of the  oscillations minimal.
\begin{figure}[tpb]
\includegraphics*[width=110mm,height=110mm]{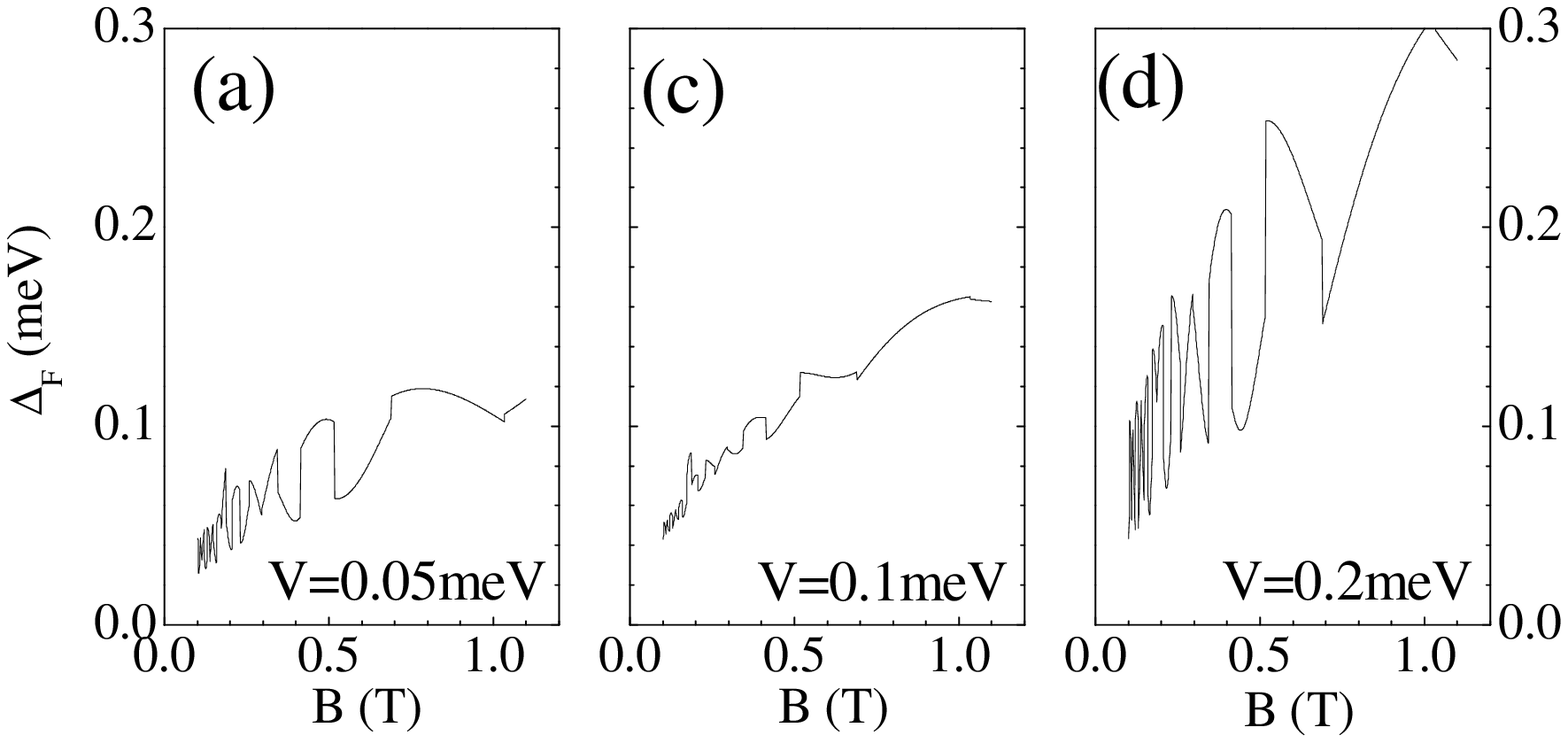}
\vspace{-2.5cm}
\caption{The same as in Fig. 7 for different electric modulation strengths $V$, as indicated, and fixed magnetic one $\hbar \omega_0=0.1meV$ at phase shift of $\pi/2$.}
\label{fig9}

\end{figure}

Finally, in Fig. 10 we plot the exact and approximate spectra of the $n=2$ Landau level,  as a function of the magnetic field, for various phase differences between the two modulations as indicated. The approximate result, for $\alpha$ integer,  is shown  by dotted curves and is again the envelope function of the exact spectrum as for  in-phase
modulations.

\begin{figure}[tpb]

\includegraphics*[width=110mm,height=150mm]{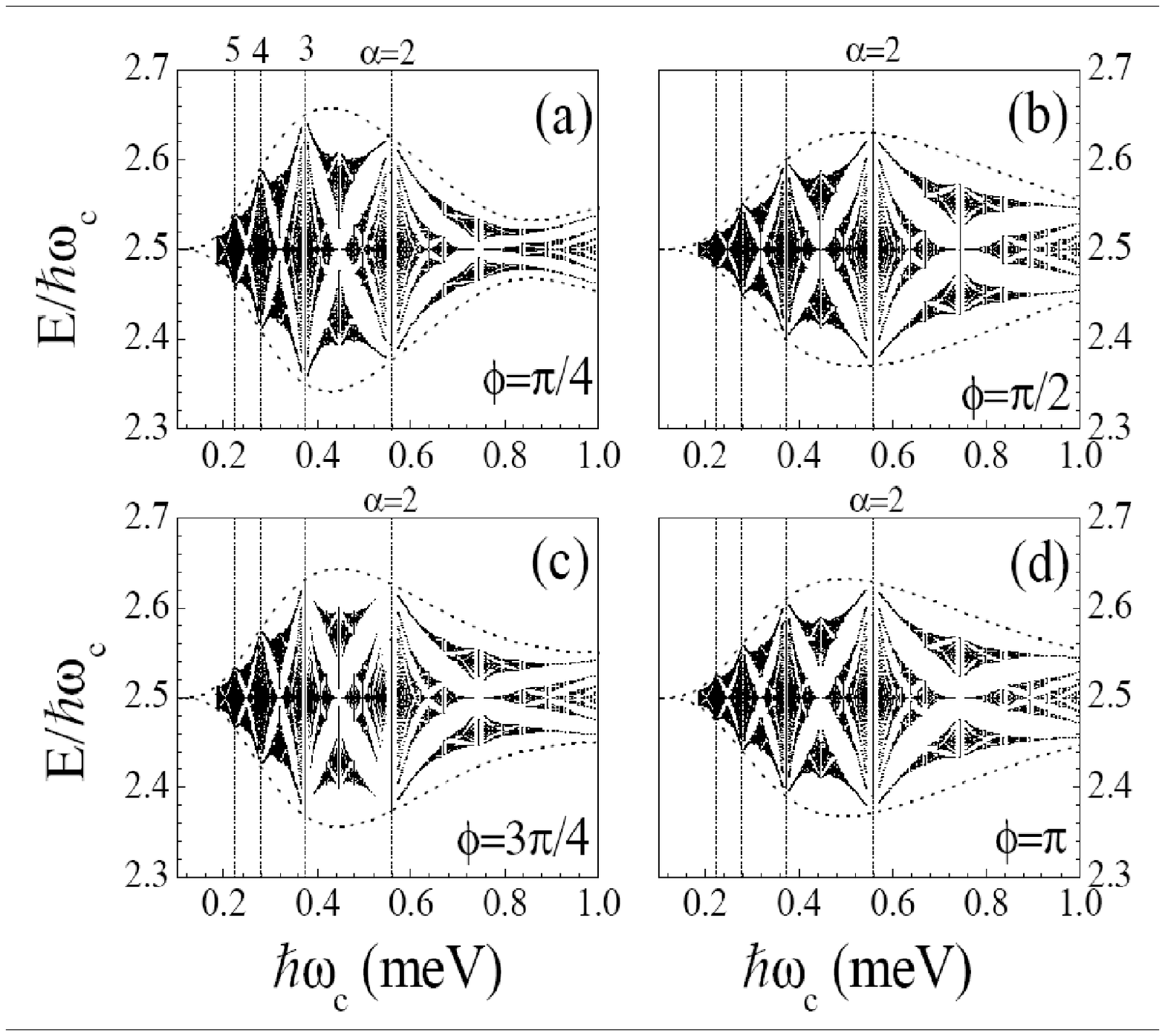}
\vspace{-2cm}
\caption{Exact and approximate (dotted curves) spectrum for various phase differences between the two modulations
$\phi=\pi/4$ (a), $\phi=\pi/2$ (b), $\phi=3\pi/4$ (c), and $\phi=\pi$ (d).}
\label{fig10}
\end{figure}

\section{Conclusions}
We studied in detail the band structure of a 2DEG in the presence of weak 2D electric and
magnetic modulations as function of an applied perpendicular magnetic field $B$. The tight-binding description shows a Hofstadter-type spectrum
with an envelope function that is determined by the strengths of  the modulations.
This envelope coincides with the
spectrum obtained from the Harper-type equation for $\alpha=\Phi_0/\Phi$ integer, where $\Phi$ is the flux per unit cell. For
$\alpha$ integer the analytical spectrum, Eq. (16),  coincides with the one obtained from  the numerical solution of Eq. (15). As an approximation, one can use the spectrum for $\alpha$ integer for all magnetic fields if the small gaps of the exact spectrum are closed due to disorder \cite{8}.
The flat-band condition and the  bandwidth are then found from the approximate bandwidth for large $n$.

For  $\alpha=q/p$ rational, the gaps in the exact spectrum, for the usual sinusoidal 2D modulations, are small and nearly close when a small level broadening is included in the calculation of the DOS
whether both modulations are present or only the electric one \cite{8}. The value of $\Gamma$ needed to close the gaps depends on the modulation strength. For instance, with $V_x=V_y=0.5$ meV, a width
$\Gamma=1.1$ K was necessary \cite{8}. In both cases, $\Gamma$ is quite small.
Accordingly, the very interesting fine structure of the spectrum may be very difficult to observe.

As discussed in the text and illustrated in Fig. 3,  the spectrum for $\alpha$ integer  and a pure electric  modulation is in general equivalent to that of a pure magnetic modulation if we neglect the very small terms  $\propto \epsilon_\mu$. The same conclusion was reached in Ref. 12 for a perurbative evaluation of the spectrum. The same analogy, though less explicitly, applies  to the exact spectrum obtained from Eq. (15) and shown in Fig.  5.

In line with our previous study \cite{8}, for other kinds of surface superlattices, e.g., hexagonal  or trigonal, the results are similar to those presented here even when we
include  cross terms $\propto V_{x}V_{y}\cos (K_{x}x)\cos (K_{y}y)$ in the modulation potential.

The oscillations of the bandwidth with the  magnetic field $B$, due to only an electric modulation, are in antiphase with those due to only a magnetic modulation.
If both modulations are present, an additional parameter is the phase between them and the situation is more complex. The  relative importance  of the two  modulations can be estimated by the parameter $\delta$, which is determined by the  modulation strengths and the electron density. The electric modulation dominates for $\delta < 1$ and the magnetic one for $\delta > 1$. By changing the electron density, the Fermi energy can be shifted from the $\delta <1 $ regime to the $\delta > 1 $ one.
Alternatively, one can change the relative strengths between the electric and magnetic modulations to tune the
value of $\delta$ which subsequently will influence the magnetoresistance.
Similar to the case of 1D  modulations \cite{13}, a transition between the two
regimes occurs for   $\delta = 1 $ and  the commensurability oscillations of
the diffusive contributions to the resistivity disappear if the two  modulations
are phase shifted by $\pi/2$.

\leftline{\bf Acknowledgments}

This work was supported by the Canadian NSERC Grant No. OGP0121756,
the Belgian Interuniversity Attraction Poles (IUAP),
the Flemish Concerted Action (GOA) Programme, and the EU-CERION programme.

\end{document}